\documentclass[sigconf]{acmart}

\usepackage{fancybox}
\usepackage{multirow}
\usepackage{flushend}
\usepackage{booktabs}
\usepackage{tabularx}
\usepackage{comment}
\usepackage{array}
\usepackage[flushleft]{threeparttable}
\usepackage{mdframed}
\graphicspath{{}{images/}{dia/}}
\DeclareGraphicsExtensions{.pdf,.png}
\usepackage{subfig}

\usepackage{listings}
\usepackage{courier}
\usepackage[normalem]{ulem}
\usepackage{color}
\usepackage{hyphenat}
\usepackage{alltt                                    % I like these
	, multirow
	,subfig
	, booktabs
	, listings
	,float
	,verbatim
	,mathtools
	,url
	,amsmath
}
\usepackage{framed,lipsum}
\usepackage[most]{tcolorbox}
\usepackage{graphics} % for EPS, load graphicx instead
\usepackage{graphicx}
\usepackage{pgfplots, pgfplotstable}
\pgfplotsset{compat=1.15}
\usepackage{tikz}
\usetikzlibrary{patterns}

\usepackage{algorithm2e}
\usepackage{algpseudocode}

\newcounter{o}
\usepackage{xspace}

\usepackage[export]{adjustbox}

\usepackage{tikz}
\usepackage{pgf-pie}
\usetikzlibrary{positioning,shadows}

\newif\ifpienumberinlegend
\pgfkeys{/number in legend/.code=
    \expandafter\let\expandafter\ifpienumberinlegend
    \csname if#1\endcsname
    \ifpienumberinlegend

    \def\beforenumber##1\afternumber{}%
    \fi,
    /number in legend/.default=true
}

\definecolor{1c1}{RGB}{188,162,6}
\definecolor{1c2}{RGB}{137,129,80}
\definecolor{1c3}{RGB}{239,167,31}
\definecolor{1c4}{RGB}{88,194,241}
\definecolor{1c5}{RGB}{6,180,188}

\tikzset{mynode/.style={draw=white,solid,circle,fill=green,inner sep=1pt, thick,
text=black}}
\tikzset{arrow line/.style={dashed, line width= 2.5pt, color=#1}}

\def\bf{\textbf}

\usepackage{paralist}

\usepackage{balance}

\normalsize

\usepackage{enumitem}

\newcommand{\nd}{\vspace{1mm}\noindent}
\usepackage{tikz}

\lstdefinestyle{inlinecode}{basicstyle={\ttfamily\scriptsize\bfseries}}

\newcommand{\urls}[1]{{\scriptsize\url{#1}}}
\usepackage{tcolorbox}

\usepackage{soul}
\usepackage{paralist}
\usepackage[outercaption]{sidecap}    

\usepackage{enumitem}
\usepackage[T1]{fontenc}
\usepackage{pifont}
\definecolor{lightgray}{gray}{.92}

\newtcolorbox
{mybox}[2][]{colbacktitle=red!10!white,
colback=blue!10!white,coltitle=black!70!black,
title={#2},fonttitle=\bfseries,#1}
\usepackage{graphicx}
\usepackage{subfig}

\definecolor{Gray}{gray}{0.9}

\copyrightyear{2021}
\acmYear{2021}
\setcopyright{acmcopyright}\acmConference[ISEC 2022]{15th Innovations in Software Engineering Conference}{February 24--26, 2022}{ DA-IICT Gandhinagar, India}
\acmPrice{15.00}
\acmDOI{XX.XXXX/XXXXXXX.XXXXXXX}
\acmISBN{XXX-X-XXXX-XXXX-X/XX/XX}

\begin{document}

% \title{Predicting Unanswered Questions of Stack Overflow During their Submission}
% \title{Early Detection of Unanswered Questions and Evidence-based Guideline to Increase Their Answerability}

\title{Reproducibility Challenges and Their Impacts on Technical Q\&A Websites: The Practitioners' Perspectives}

\author{Saikat Mondal}
\affiliation{%
  \institution{University of Saskatchewan, Canada}
%   \city{Saskatoon}
%   \country{Canada}
    }
\email{saikat.mondal@usask.ca}

\author{Banani Roy}
\affiliation{%
  \institution{University of Saskatchewan, Canada}
%   \city{Saskatoon}
%   \country{Canada}
  }
\email{banani.roy@usask.ca}

% \affiliation{%

%PredUnations

% \author{\IEEEauthorblockN {Saikat Mondal\hspace{0.7cm}Avijit Bhattacharjee\hspace{0.7cm}Chanchal K. Roy}
% \IEEEauthorblockA{SRlab, Department of Computer Science, University of Saskatchewan\\ }}
% \ saikat.mondal@usask.ca}
% }

\renewcommand{\shortauthors}{Saikat Mondal et al.}
\renewcommand{\shorttitle}{Reproducibility Challenges and Their Impacts on Technical Q\&A Websites}

\begin{abstract}
Software developers often look for solutions to their code-level problems by submitting questions to technical Q\&A websites like Stack Overflow (SO). They usually include example code segments with questions to describe the programming issues. SO users prefer to reproduce the reported issues using the given code segments when they attempt to answer the questions. Unfortunately, such code segments could not always reproduce the issues due to several unmet challenges (e.g., external library not found) that might prevent questions from receiving prompt and appropriate solutions. A previous study produced a catalog of potential challenges that hinder the reproducibility of issues reported at SO questions. However, it is unknown how the practitioners (i.e., developers) perceive the challenge catalog. Understanding the developers' perspective is inevitable to introduce interactive tool support that promotes reproducibility. We thus attempt to understand developers' perspectives by surveying 53 users of SO. In particular, we attempt to -- (1) see developers' viewpoints on the agreement to those challenges, (2) find the potential impact of those challenges, (3) see how developers address them, and (4) determine and prioritize tool support needs. Survey results show that about 90\% of participants agree to the already exposed challenges. However, they report some additional challenges (e.g., error log missing) that might prevent reproducibility. According to the participants, too short code segment and absence of required Class/Interface/Method from code segments severely prevent reproducibility, followed by missing important part of code. To promote reproducibility, participants strongly recommend introducing tool support that interacts with question submitters with suggestions for improving the code segments if the given code segments fail to reproduce the issues.
\end{abstract}



%%\keywords{Theory, Metrics, Human factors}

% \setcopyright{acmcopyright}
% \copyrightyear{2021}
% \acmYear{2021}
% \acmDOI{10.1145/3452383.3452392}

% %% These commands are for a PROCEEDINGS abstract or paper.
% \acmConference[ISEC 2021]{ISEC '21: Innovations in Software Engineering Conference}{February 25-27, 2021}{Bhubaneswar, India}
% % \acmBooktitle{ISEC 2021: Innovations in Software Engineering Conference,
% %   February 25-27, 2021, Bhubaneswar, India}
% \acmPrice{15.00}
% \acmISBN{978-1-4503-9046-0/21/02}

\maketitle
\section{Introduction}
\label{introduction}
Stack Overflow (SO) has emerged as one of the largest and most popular technical question and answer (Q\&A) sites. SO is continuously contributing to the body of knowledge in software development \citep{tahaei2020understanding, treude2011programmers, vincent2018examining}. Millions of software developers interact through SO every month to solve their programming-related problems \citep{gao2020code2que}. About seven thousand questions are submitted in SO every day. Among them, a large number of questions discuss code level problems (e.g., errors, unexpected behaviour) \citep{treude2011programmers}. Such questions often include problematic code segments with the programming issue descriptions. SO users generally prefer to reproduce the issues reported at the questions using the code segments and then submit their solutions \citep{Mondal-SOIssueReproducability-MSR2019}. Reproducibility means a complete agreement between the reported and investigated issues \citep{crashdroid, Mondal-SOIssueReproducability-MSR2019}. Unfortunately, such programming issues could not always be reproduced by other users due to several unmet challenges (e.g., too short code segment) of the code segments \citep{Mondal-SOIssueReproducability-MSR2019, crashscope, soltani2017guided}. This phenomenon prevents the questions from getting prompt and appropriate solutions. Mondal et al.~ \citep{Mondal-SOIssueReproducability-MSR2019} show that a question whose code segment could reproduce the reported issue has more than three times higher chance of receiving an acceptable answer (i.e., working solution) than the question whose code segment could not reproduce the issue. Moreover, the median time delay of receiving an accepted answer is double, and the average number of answers is significantly less for the questions with irreproducible issues than those of the questions with reproducible issues.
%Thus, a detailed investigation is warranted on the challenges that prevent the reproducibility of issues reported at SO questions.

\begin{table}[htb]
	\centering
	\caption{Challenges preventing issue reproduction}
	\label{table:reproducibility-challenges}
    \resizebox{3.2in}{!}{%
    %\rowcolors{1}{}{lightgray}
    \begin{tabular}{p{8cm}}  \toprule

    $\bullet$ Class/Interface/Method not found $\bullet$ Important part of code missing $\bullet$ External library not found $\bullet$ Identifier/Object type not found $\bullet$ Too short code snippet $\bullet$ Database/File/UI dependency $\bullet$ Outdated code  \\ \bottomrule
    \end{tabular}
    }
\end{table}

A couple of existing studies \citep{querytousablecode, gistable} investigate the challenges of usability and executability of the code segments posted on Q\&A sites. For example, a study conducted by Yang et al.~\citep{querytousablecode} analyzes the parsability and compilability of code segments extracted from the accepted answers of SO. However, their analysis was fully automatic that only exposes parse and compile errors (e.g., syntax errors,  incompatible types). Horton and Parnin~\citep{gistable} examine the executability of the Python code segments found on the GitHub Gist system. They identify several flaws (e.g., syntax errors, indentation errors) that prevent the executability of such code segments. However, a simple execution success does not always guarantee the reproducibility of issues \citep{Mondal-SOIssueReproducability-MSR2019}. Several studies investigate the quality of SO code segments by measuring their readability \citep{buse2008metric, buse2009learning, daka2015modeling, posnett2011simpler, scalabrino2016improving, treude2017understanding}, and understandability \citep{lin2008evaluation, scalabrino2017automatically, trockman2018automatically}. Unfortunately, their capability of reproducing the issues reported at SO questions was not investigated. Thus, their approach also fails to address reproducibility challenges. 
% Mu et al.~\citep{mu2018understanding} investigate the crowd-reported security vulnerabilities to quantify their reproducibility. They later survey hackers, researchers, and engineers who have domain expertise in software security to understand their viewpoint on the prevalence of missing information in vulnerability reports. While our work overlaps with them in terms of methodology, our research goal and context are different. 
Mondal et al.~\citep{Mondal-SOIssueReproducability-MSR2019} first investigate the reproducibility of issues reported at SO questions related to Java programming language. Their investigation produces a catalog of challenges (see Table \ref{table:reproducibility-challenges}) that might prevent reproducibility. However, the catalog was not validated by the practitioners (i.e., developers). Thus, it is unknown to us how the developers perceive the reproducibility challenges. Understanding the developers' perspective is important to introduce efficient tool support to promote reproducibility.

This study addresses developers' perspectives on the reproducibility challenges and their impact by surveying 53 software developers. In particular, we attempt to -- (1) validate the challenge catalog (see Table \ref{table:reproducibility-challenges}) by asking developers' agreement to those challenges, (2) understand the potential impact of each of those challenges to answer SO questions, (3) see how developers address reproducibility challenges, and (4) find developers' recommendations on designing efficient tool support to help question submitters improve their code segments to promote reproducibility.

\nd\bf{Replication Package} that contains the survey questionnaire and all the responses is shared in our online appendix \citep{ourdataset}.

\section{Background and Related Work}
\label{sec:background}
In this section, we start by introducing the issue reproducibility and the challenges that prevent reproducibility. We then proceed to describe related studies. 

\begin{figure}[!htb]
\centering
  \includegraphics[width=2.5in]{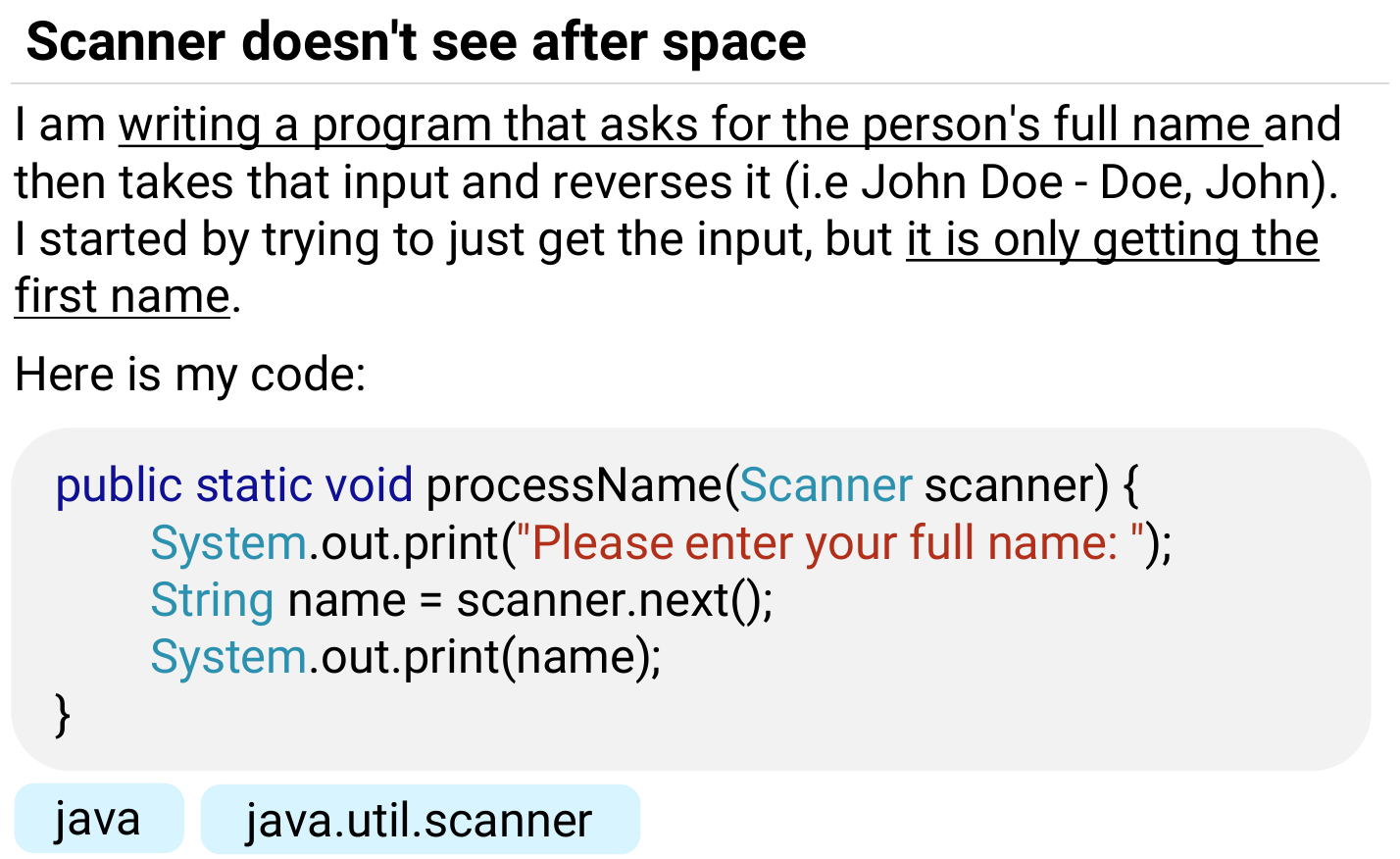}
  \caption{\small{An example \citep{footnote1} question of SO that discusses a programming issue.}}
  \label{fig:example-issue-reproducibility}
\end{figure}
%\footnotetext[1]{https://stackoverflow.com/questions/19509647}

\subsection{Issue Reproducibility}
Reproducibility is often closely related to repeatability and replicability \citep{anda2008variability}. However, its definition differs across disciplines. In this study, reproducibility means a complete agreement between the reported and the investigated issues \citep{crashdroid, Mondal-SOIssueReproducability-MSR2019}. Consider the example question in Fig. \ref{fig:example-issue-reproducibility}, where a user was trying to take a person's full name (e.g., John Doe) as input by invoking the \texttt{next()} method of Java \texttt{Scanner} class. However, when the user was printing the name, only the first name (e.g., John) was being printed, and the last name (e.g., Doe) was getting lost. In particular, the user was not getting the part of the name after space. She included the definition of the method \texttt{processName()} with the question description, where she was taking and printing the name. If other users also find the first part of the name and lose the part after space by invoking \texttt{processName()} method, that means the issue is reproducible. On the contrary, when others fail to regenerate the reported issue using the given method, it suggests the issue is not reproducible.

\subsection{Reproducibility Challenges}

Mondal et al.~\citep{Mondal-SOIssueReproducability-MSR2019} analyze 400 SO questions related to the Java programming language and attempt to reproduce their issues. However, they could not reproduce about 22\% issues due to several unmet challenges of the code segments. Table \ref{table:reproducibility-challenges} shows those reproducibility challenges. In this section, we discuss the challenges that prevent reproducibility as follows.

\begin{figure}[htb]
\centering
  \includegraphics[width=2.5in]{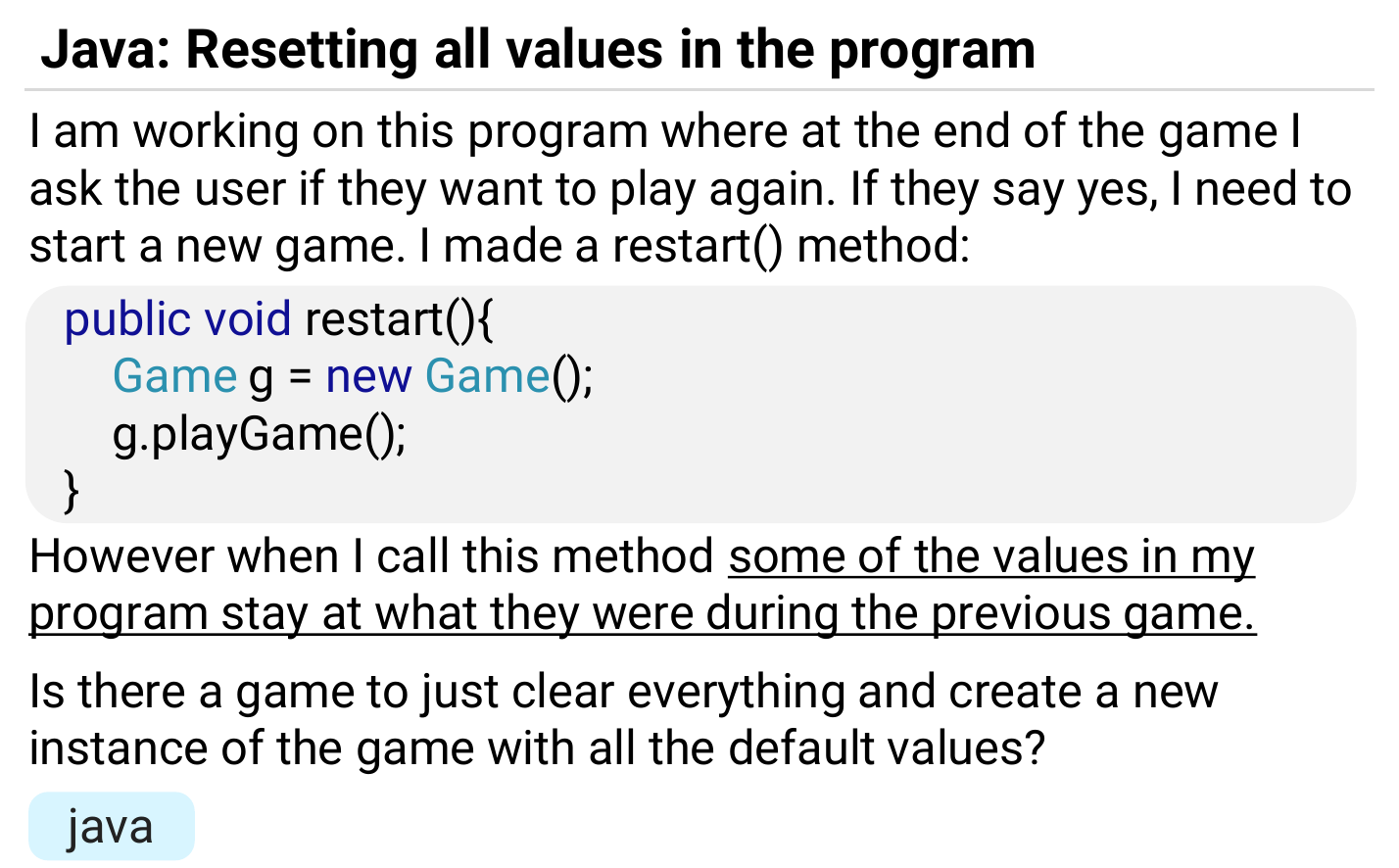}
  \caption{\small{An example \citep{footnote2} question of SO whose issue could not be reproduced due to mainly two unmet challenges -- (i) class/interface/method not found and (ii) important part of code missing.}}
  \label{fig:reproducibility-challenge-1}
\end{figure}
%\footnotetext[2]{https://stackoverflow.com/questions/798184}

\begin{inparaenum}[(1)]

    \item \textbf{Class/Interface/Method not found.} Developers often submit only the code segments of interest with questions, which are neither complete nor compilable. Such code segments also invoke methods from classes. However, the code segments often miss the definition of the methods. Let us consider the example question in Fig. \ref{fig:reproducibility-challenge-1}, where the developer attempts to reset all the variables by the default values. However, the code was not working as expected, and some of the variables were retaining their old values. Unfortunately, this issue could not be reproduced due to several unmet challenges. Especially, definitions of the \texttt{Game} class and \texttt{playGame()} method are missed but essential to reproduce the issue. Thus, one developer commented while attempting to answer the question – \emph{``Can you post more code? The Game class? The class that contains the restart() method?''}.
    
    \item \textbf{Important part of code missing.} Code segments included with questions often miss such statements (i.e., part of code) without which issues could not be reproduced. Here, ``an important part'' means such part of code that could never be guessed properly. For instance, we could add the definition of \texttt{Game} class and \texttt{playGame()} method in the code example in Fig. \ref{fig:reproducibility-challenge-1} and make the code executable. However, we cannot reproduce the issue. We need the part of the code inside the class and the method by which the developer attempted to reset the variables.
    
\begin{figure}[htb]
  \centering
  \includegraphics[width=2.5in]{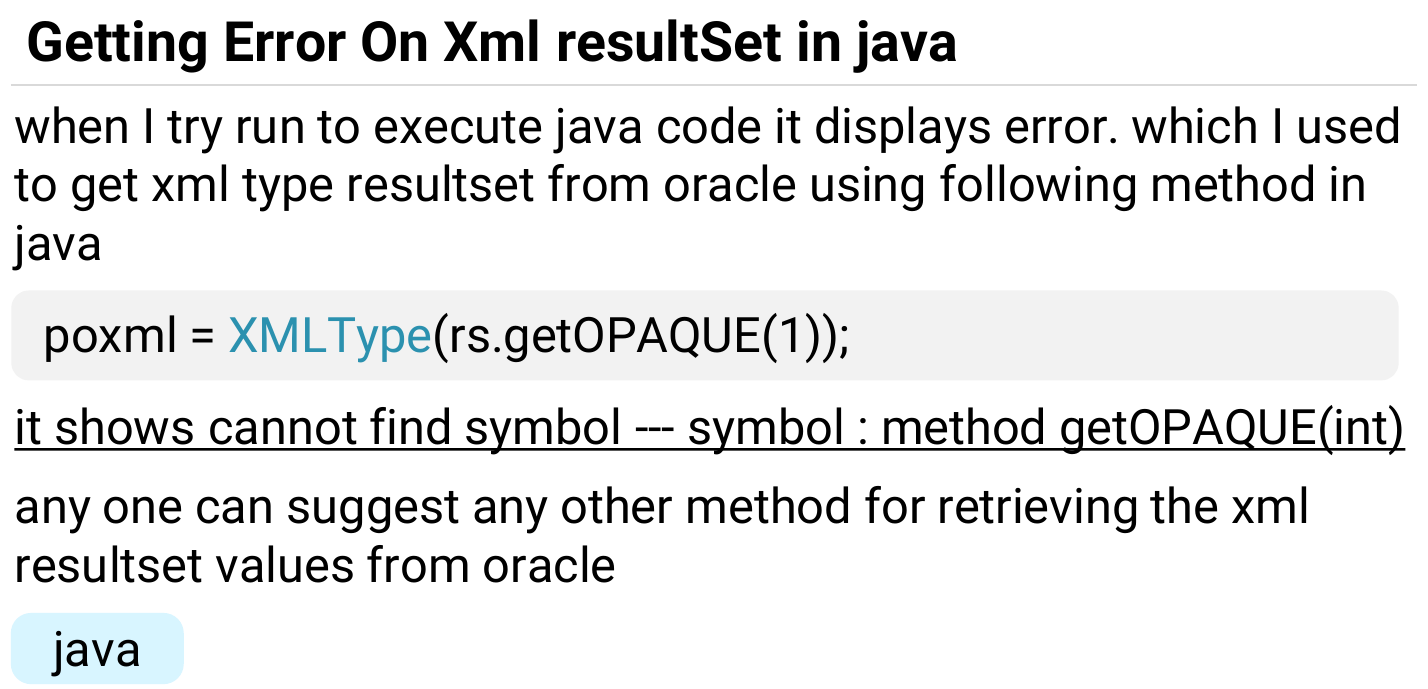}
  \caption{\small{An example \citep{footnote3} question of SO whose issue could not be reproduced due to mainly three unmet challenges -- (i) external library not found, (ii) identifier/object type not found and (iii) too short code snippet.}}
  \label{fig:reproducibility-challenge-2}
\end{figure}
%\footnotetext[3]{https://stackoverflow.com/questions/2018247}
    
    \item \textbf{External library not found.} Resolving external library dependencies is one of the major challenges to reproduce the issues from the submitted code segments \citep{Mondal-SOIssueReproducability-MSR2019}. Java has thousands of external (i.e., third-party) libraries with millions of classes and methods. However, different libraries contain classes and methods with the same name. Thus, if the code segments miss import statements of the external libraries or question descriptions do not have hints that point to the appropriate libraries, developers face difficulty adding the appropriate libraries. As a result, they could not compile/execute the code and fail to reproduce the question issues. Consider the code example shown in Fig. \ref{fig:reproducibility-challenge-2}, where the import statement for the library that contains class \texttt{XMLType} was missing. Thus, the inclusion of the appropriate external library is a major challenge here to reproduce the question issue. 
    
    \item \textbf{Identifier/Object type not found.} Code segments use identifiers/objects without declaring them. In some cases, developers infer the type of identifiers/objects by looking at the assigned values or method invocations. Otherwise, they find difficulties in inferring their types that might prevent the reproducibility of issues. Consider the code example in Fig. \ref{fig:reproducibility-challenge-2}, where the types of \texttt{poxml} and \texttt{rs} are unknown and hard to guess from the submitted code.
    
    \item \textbf{Too short code snippet.} Developers often submit incomplete code segments with their questions. However, code segments are too short to reproduce the question issues in many cases. For example, Fig. \ref{fig:reproducibility-challenge-2} shows a question where the question submitter included only one line of code with the issue description. Thus, it is genuinely challenging to guess the missing statements and make the code compilable/executable to reproduce the issue. The challenge ``too short code segment'' could overlap with an ``important part of code missing''. However, an important part of code could be missed even when the developers submit a long code.
    
    \begin{figure}[htb]
      \centering
      \includegraphics[width=2.8in]{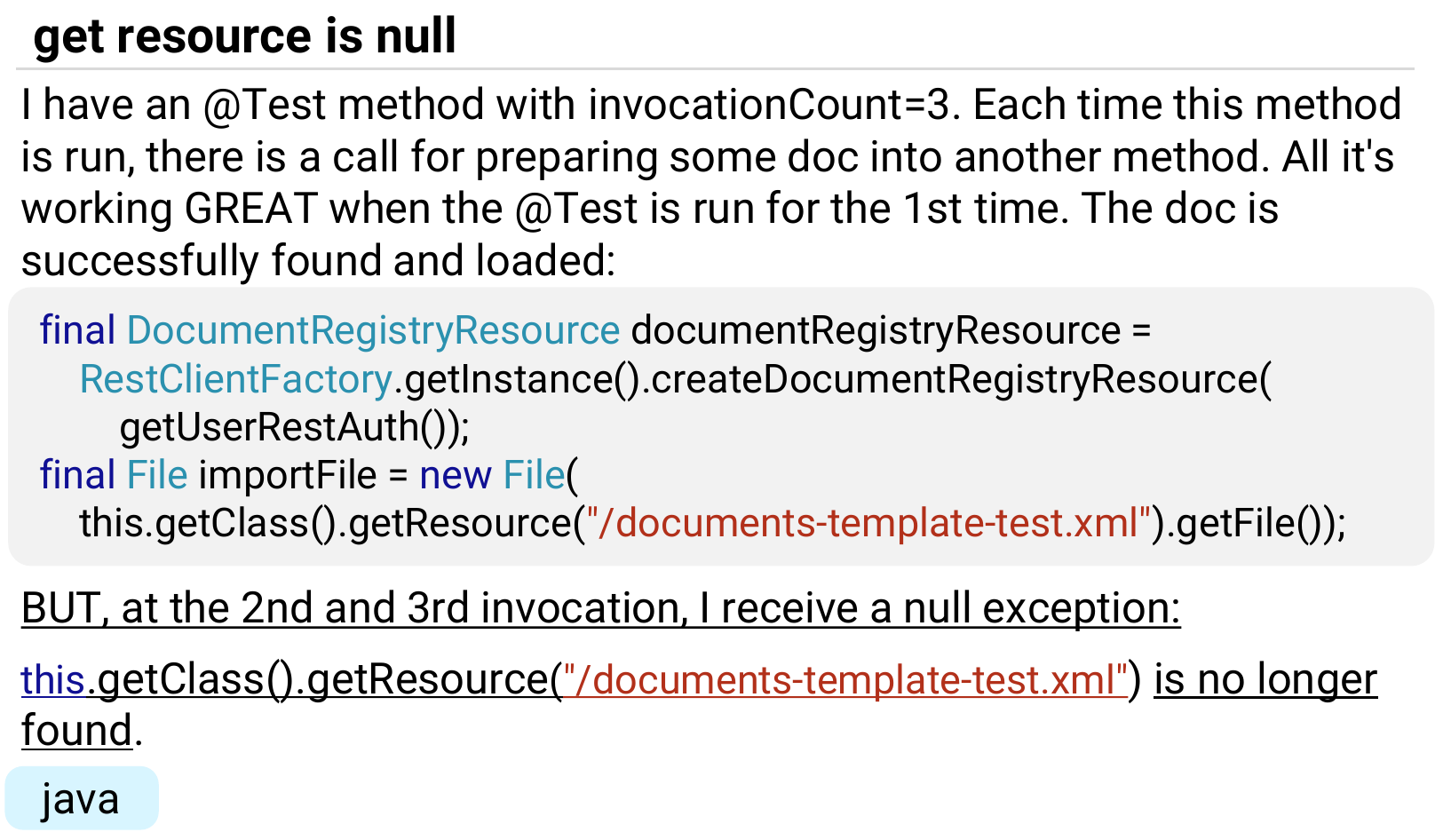}
      \caption{\small{An example \citep{footnote4} question of SO whose issue could not be reproduced due to mainly two unmet challenges -- (i) database/file/UI dependency and (ii) class/interface/method not found.}}
      \label{fig:reproducibility-challenge-3}
    \end{figure}
    %\footnotetext[4]{https://stackoverflow.com/questions/2012643}
        
    \begin{figure}[htb]
      \centering
      \includegraphics[width=3in]{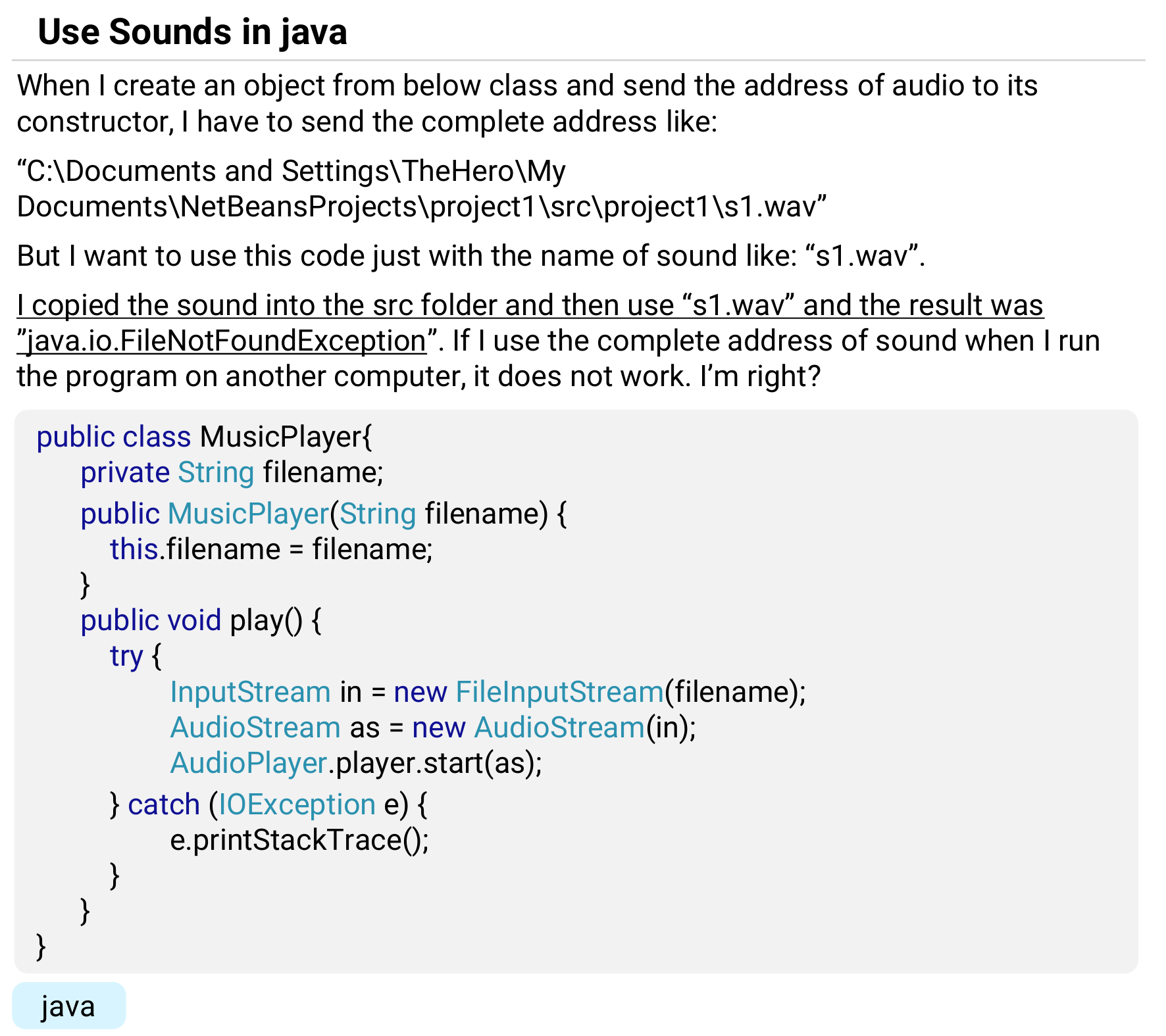}
      \caption{\small{An example \citep{footnote5} question of SO whose issue could not be reproduced due to mainly outdated code challenge.}}
      \label{fig:reproducibility-challenge-4}
    \end{figure}
    %\footnotetext[5]{https://stackoverflow.com/questions/1264770}
    
    \item \textbf{Database/File/UI dependency.} Several code segments could not reproduce the issues due to their complex interactions with databases, external files, and UI elements. Consider the example question in Fig. \ref{fig:reproducibility-challenge-3}, where the developer attempted to create and load an XML doc. Unfortunately, the code got a null exception when the developer attempted to access the doc more than once. Such an issue could not be reproduced due to the external file dependency. However, the doc could not be created because the question lacks important detail (e.g., file path, content) associated with the file.
    
    \item \textbf{Outdated code.} A few code segments contain outdated code (e.g., deprecated class/API) that could prevent issue reproducibility. A few studies investigate the outdated code of Stack Overflow \citep{zhang2019empirical,ragkhitwetsagul2019toxic}. However, in this study, we consider a code is outdated when the code contains such classes/APIs that are not compatible with JDK-1.8. It is hard to find an equivalent or alternative class/API for the outdated class/API in some cases. Consider the code segment in Fig. \ref{fig:reproducibility-challenge-4}, where the classes \texttt{AudioStream} and \texttt{AudioPlayer} are not being used in JDK-1.8. Thus, the developers face difficulties in finding their alternatives while reproducing the question issue.

\end{inparaenum}

\subsection{Related Work}
\label{relatedwork}

Several studies investigate usability (e.g., parsability, compilability) \citep{querytousablecode}, executability \citep{gistable} and reproducibility  \citep{querytousablecode, gistable, mu2018understanding, rahman2020why, Mondal-SOIssueReproducability-MSR2019, crashdroid,crashscope,soltani2017guided,yakusu} challenges of the code segments posted at crowd-sourced developer forums (e.g., GitHub, SO). However, our study first investigates the practitioners' perspective on the reproducibility challenges of the programming issues reported at SO questions, their impact to answer questions, and interactive tool design requirements to promote reproducibility.

Yang et al.~\citep{querytousablecode} analyze the usability of about 914K Java code segments extracted from the accepted answers of SO. They expose several challenges of their parsability (e.g., syntax error) and compilability (e.g., incompatible types). The authors employ automated tools such as Eclipse JDT and ASTParser to parse and compile the code segments and report the challenges that prevent them from parsing and compiling. However, they do not analyze the challenges of reproducibility. Horton and Parnin~\citep{gistable} investigate the executability of Python code found on the GitHub Gist system. They report the types of execution failures encountered while running Python gists such as import error, syntax error, indentation error. However, a code segment's execution success does not always guarantee the reproducibility of an issue reported at the SO question. Reproducibility may require testing and debugging that warrant manual analysis, which was not done by Horton and Parnin~\citep{gistable}. Mondal et al. \citep{Mondal-SOIssueReproducability-MSR2019} go beyond code execution and manually investigate the reproducibility of issues reported at 400 questions related to Java programming language using the code segments included with questions. They produce a catalog of challenges that prevent reproducibility of SO question issues. However, it is important to listen to the practitioners whether they agree to the challenges and understand the potential impact of those challenges to answer a question that was not done by any earlier studies.

Several researchers investigate the reproducibility challenges of software bugs and security vulnerabilities \citep{erfani2014works, rahman2020why, mu2018understanding}. Joorabchi et al. \citep{erfani2014works} analyze 1,643 irreproducible bug reports and investigate the causes of their irreproducibility. They reveal six root causes, such as environmental differences, insufficient information that prevent bug-reports' reproducibility. Rahman et al.~\citep{rahman2020why} conduct a study to understand the irreproducibility of software bugs. They investigate 576 irreproducible bug reports from two popular software systems (e.g., Firefox, Eclipse) and identify 11 challenges (e.g., bug duplication, missing information, ambiguous specifications) that might prevent bug reproducibility. The authors then survey 13 developers to understand how the developers cope with irreproducible bugs. According to the study findings, developers either close these bugs or solicit further information. Mu et al.~\citep{mu2018understanding} analyze 368 security vulnerabilities to quantify their reproducibility. Their study suggests that individual vulnerability reports are insufficient to reproduce the reported vulnerabilities due to missing information. Besides, many vulnerability reports do not include details of software installation options and configurations, or the affected operating system (OS) that could hinder reproducibility. Then they survey hackers, researchers, and engineers who have domain expertise in software security. Survey findings suggest that apart from internet-scale crowd-sourcing and some interesting heuristics, manual efforts (e.g., debugging) based on experience are the sole way to retrieve missing information from reports. Our study relates to the above studies in terms of research methodologies and problem aspects. However, our research context differs from theirs since we attempt to understand the practitioners' perspectives on SO questions' reproducibility issues.

Tahaei et al.~\citep{tahaei2020understanding} analyze 1,733 privacy-related questions of SO to understand the challenges and confusion that developers face while dealing with privacy-related topics. Ebert et al.~\citep{ebert2019confusion} investigate the reasons (e.g., missing rationale) and impacts (e.g., merge decision is delayed) of confusion in code reviews. Their study suggests how developers cope with confusion during code reviews. For example, developers attempt to deal with confusion by requesting information, improving the familiarity with existing code, and discussing off-line the code review tool. They survey developers to obtain actionable insights for both researchers and tool builders. While our work overlaps with them in terms of methodology, however, our research goals are different. They attempt to find reasons and impacts of confusion in code reviews. However, we survey 53 developers to understand (1) the impacts of reproducibility of SO questions issues, (2) how developers plan to modify the code segments to make them capable of reproducing the issues, and (3) the insights for introducing interactive tool supports.

Ford et al.~\citep{ford2018we} deploy a month-long, just-in-time mentorship program to SO to guide the novices with formative feedback of their questions. Such mentorship reduces the negative experience caused by delays in getting answers or adverse feedback. However, human mentorship is costly. Thus, it is hard to sustain such mentorship. Horton and Parnin~\citep{horton2019dockerizeme} present DockerizeMe, a system for inferring the dependencies required to execute a Python code snippet without import errors. Terragni et al.~\citep{terragni2016csnippex} propose a technique CSnippEx to automatically convert Java code segments into compilable Java source code files. 
However, we plan to determine the design requirement to introduce an intelligent and interactive tool based on the developers' recommendations. Such a tool could analyze SO code segments included with questions related to Java programming problems and suggest users improve the code segments to promote reproducibility.

\section{Study Design}

\subsection{Objective and Research Questions}

In this study, we aim to understand developers' perspectives on reproducibility challenges and estimate their impacts to answer SO questions. We also plan to introduce tool support to promote reproducibility by seeking developers' suggestions on the design of the tool. Following this aim, we guide our study with five research questions and make five contributions in this paper as follows.

\vspace{1mm}
$\bullet$ \textbf{(RQ\textsubscript{1}) What do developers consider to be the challenges behind the irreproducibility of issues reported at Stack Overflow questions?}
Mondal et al.~\citep{Mondal-SOIssueReproducability-MSR2019} produce a catalog of challenges (see Table \ref{table:reproducibility-challenges}) that might prevent reproducibility. However, developers' feedback on such empirical findings is essential to increase confidence in the findings \citep{rahman2020why}. 
To answer RQ\textsubscript{1}, we present four example questions of SO with their issue descriptions and code segments. Participants were asked to reproduce the reported issues using the code segments. Upon failure, we ask for their agreement to the given reproducibility challenges. The given challenges were validated by 53 software developers with an agreement level between 80\% and 94\%. 

%\vspace{1mm}
$\bullet$ \textbf{(RQ\textsubscript{2}) What are the perceived impacts of the reproducibility challenges to answer a Stack Overflow question?}
Understanding the impact of each of the challenges is important to determine which challenges must be resolved by question submitters to receive appropriate solutions to their questions. To answer RQ\textsubscript{2}, we present each of the challenges with five options -- (1) \emph{not a problem}, (2) \emph{moderate}, (3) \emph{severe}, (4) \emph{blocker}, and (5) \emph{no opinion}. We then ask participants to select one of the five options. Participants assessed ``an important part of code missing'' as mostly a blocker, and ``outdated code'' as mostly not a problem.

%\vspace{1mm}
$\bullet$ \textbf{(RQ\textsubscript{3}) How do developers prioritize to address the reproducibility challenges?}
Prioritization of challenges is important to determine which challenges need to be fixed first within a strict budget (e.g., time constraint). We thus ask participants to specify three reproducibility challenges they would like to prioritize above others. We find the following challenges in the top three list -- (1) an important part of code missing, (2) class/interface/method not found, and (3) too short code snippet. 

%\vspace{1mm}
$\bullet$ \textbf{(RQ\textsubscript{4}) How do developers plan to modify the code segments to make them capable of reproducing the question issues?}
Developers' code editing plan to reproduce the question issues offers more insights into how an intelligent tool could suggest question submitters improve their code segments. We thus attempt to observe participants' editing actions to address the reproducibility challenges. We see that participants perform several editing actions such as adding demo classes and methods, declaring and initializing variables, invoking methods, including external and native libraries. In particular, the participants first attempt to make the code segments compilable/executable to reproduce the issues. On the contrary, several participants do not modify the code segments. They consider the code segments insufficient to make them compilable/executable, or modifications based on assumptions (without appropriate hints) could deviate the code segments much from the original code.

%\vspace{1mm}
$\bullet$ \textbf{(RQ\textsubscript{5}) What are the interactive tool design requirements to address the reproducibility challenges?}
Questions with reproducible issues have a significantly higher chance of receiving acceptable answers with minimum time delay \citep{Mondal-SOIssueReproducability-MSR2019}. However, the current question submission system of SO is not capable of addressing reproducibility issues. To introduce tool support, we thus ask participants' recommendations on tool design to promote reproducibility. According to them, intelligent tools that (1) analyze the code segments statically to find the reproducibility challenges and (2) suggest question submitters improving the code examples could help them receiving appropriate solutions to their questions.

\subsection{Methodology}

In the following, we describe the study methodology. First, we survey to understand \emph{``what developers say''} (\textbf{RQ\textsubscript{1}}, \textbf{RQ\textsubscript{2}}, and \textbf{RQ\textsubscript{3}}) about reproducibility challenges and their potential impact to answer questions. Then we see the code modification plan to understand \emph{``what developers do''} (\textbf{RQ\textsubscript{4}}) to reproduce the issues. Next, we find \emph{``what developers suggest''} (\textbf{RQ\textsubscript{5}}) to receive recommendations that inform the design needs of the interactive tool supports to assist reproducibility of question issues. Finally, we analyze the survey findings and report them.

\subsubsection{Survey}
We conduct an online survey aiming at validating and extending the catalog of reproducibility challenges (see Table \ref{table:reproducibility-challenges}) identified by Mondal et al.~\citep{Mondal-SOIssueReproducability-MSR2019}. Furthermore, we ask the participants to give their viewpoints on the impacts of those challenges and their potential fixes. We primarily follow Kitchenham and Pfleeger's guidelines for personal opinion surveys \citep{kitchenham2008personal}. However, we also consider the guidance and ethical issues from the established best practices~\citep{groves2011survey, singer2002ethical}.

% \vspace{2mm}
\textbf{Survey Design.}
Our survey includes different types of questions (e.g., multiple-choice and free-text answers). Before asking questions, we explain the purpose of the survey and our research goals to the participants. We ensure the survey participants that the information they provide must be treated confidentially. We first piloted the preliminary survey with a small set of practitioners (i.e., 3 participants). We collect their feedback on (1) whether the length of the survey was appropriate and (2) the clarity and understandability of the terms. We then perform minor modifications to the survey draft based on the received feedback and produce a final version. We inform the estimated time (i.e., 45--50 minutes) required to complete the survey to the participants based on the pilot survey. We exclude the three responses from the pilot survey from the presented results in this paper. 
Our survey comprises five parts as follows.

\begin{inparaenum}[(i)]%[\bfseries ({A}1)]

    \item\emph{Consent and Prerequisite.} In this part, we ask the participants to confirm whether they consent to participate in this survey and agree to process their data. We also ask questions to confirm whether they answer SO questions and have experience in Object-Oriented Programming (OOP), especially Java. Otherwise, we did not allow them to participate in our survey.
    
    \item\emph{Participants Information.} In this part, we attempt to collect some information about the participants, such as years of experience in OOP and professional software development, their current profession, SO account age, and their question answering experience in SO.
    
    \item\emph{Agreement to the Reproducibility Challenges.} In this part, we present four SO questions from the manually analyzed dataset by Mondal et al. \citep{Mondal-SOIssueReproducability-MSR2019}. We ask participants to reproduce the question issues using the code segments. Upon failure, we ask their agreement/disagreement with the challenges. Otherwise, we ask for their code editing actions required to perform to reproduce the issues. In both cases, there are options to report additional challenges or editing actions.
    
    \item\emph{Impact of the Reproducibility Challenges.} Here, we ask questions to understand the potential impact and severity of each of the reproducibility challenges.
    
    \item\emph{Tool Support Needs.} Finally, we seek the participants' opinions on the needs and design requirements of an intelligent tool to promote reproducibility. We also offer a few tool support options and employing a 5-point Likert Scale \citep{joshi2015likert, vagias2006likert} to estimate the participants' responses.
    
\end{inparaenum}

\begin{figure}[t]
	\centering
	\resizebox{3.4in}{!}{
	\subfloat[Java experience (in years)]
    {
	\resizebox{3in}{!}{
    \begin{tikzpicture}
    %   \pie [polar, explode=0.1, color={black!5, black!15, black!25}, text=legend]
    \pie[explode=0.1, text=pin, number in legend, sum = auto, color={black!0, black!15, black!30, black!45, black!60, black!75}]
        { 23/\Huge{\textbf{$\leq$ 2} (43.4\%)},
          11/\Huge{3--5 (20.7\%)},
          8/\Huge{6--8 (15.1\%)},
          6/\Huge{9--11 (11.3\%)},
          2/\Huge{12--14 (3.8\%)},
          3/\Huge{$\geq$ 15 (5.7\%)}
        }
    \end{tikzpicture}
    \label{fig:java-experience}
    }
    }
    \subfloat[Profession]
    {
	\resizebox{2.4in}{!}{
    \begin{tikzpicture}
    %   \pie [polar, explode=0.1, color={black!5, black!15, black!25}, text=legend]
    \pie[explode=0.1, text=pin, number in legend, sum = auto, color={black!0, black!20, black!40, black!60}]
        { 24/\Huge{SD (45.3\%)},
          4/\Huge{TL (7.6\%)},
          5/\Huge{RE (9.4\%)},
          20/\Huge{AP (37.7\%)}
          }
    \end{tikzpicture}
    \label{fig:profession}
    }
    }
    }
\caption{Information of the survey participants \footnotesize{(\textbf{SD:} Software Developer, \textbf{TL:} Technical Lead, \textbf{RE:} Research Engineer, \textbf{AP:} Academic Practitioner)}.}
\label{fig:info-survey-participants}
\vspace{-4mm}
\end{figure}
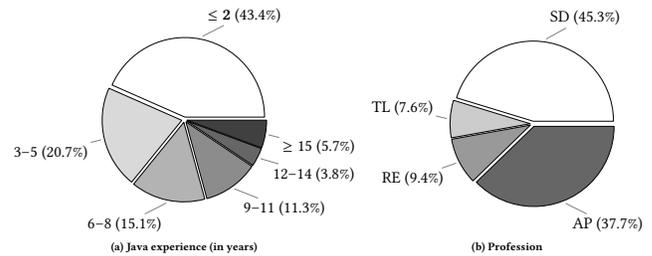

% \vspace{2mm}
\textbf{Recruitment of Survey Participants.} We recruit participants in the following two ways.

\begin{inparaenum}[(i)]%[\bfseries ({A}1)]

    \item \emph{Snowball Approach:} We recruit a list of participants from a set of companies worldwide based on personal contacts. We then adopt a snowballing method \citep{bi2021accessibility} to encourage the participants to disseminate our survey to some of their colleagues with similar experiences and willing to participate in our survey. In this process, we confirm 48 participants. However, we allow 44 of them to complete our survey. The remaining four did not satisfy our constraints perfectly.
    
    \item \emph{Open Circular:} To find potential participants, we post a description of this study and our research goals in the specialized Facebook groups where professional software developers discuss their programming problems and share software development resources. We also use LinkedIn as a research tool to reach potential participants because it is one of the largest professional social networks in the world. We get 15 participants from this open circular who are willing to participate and satisfy our constraints. Finally, we got nine valid responses from them.

\end{inparaenum}

Fig. \ref{fig:info-survey-participants} summarizes the participants' Java experience and professions. Out of the 53 participants, 34 (about 64\%) had Java experience five years or less (see Fig. \ref{fig:java-experience}). Developers with comparatively less experience are likely to be more actively engaged in Stack Overflow. However, we recruit 19 participants who had more than five years of Java development experience. The professions of the survey participants were mainly software developers (45.3\%), and academic practitioners (e.g., faculty member, student, postdoctoral researcher) (37.7\%) (see \ref{fig:profession}). However, there were four technical leads and five research engineers. Other details of the participants (e.g., professional software development experience, SO account age, and question answering experience) can be found in our online appendix \citep{ourdataset}.

% \begin{itemize}
%     \item \textbf{Snowball Approach.} We adopt a snowball approach to recruit our study participants. First, we recruit a list of participants who meet our recruitment criteria based on personal contacts. We then ask the participants to recommend other developers with similar experiences. In this process, we confirm 48 participants. However, four of their responses were incomplete. We thus discard them and accept 44 valid responses. 
    
%     \item \textbf{Open Circular.} To find potential participants, we posted a description of this study and our research goals in Facebook\footnotemark[6]\footnotetext[6]{https://www.facebook.com/}, the most popular social interaction site. We primarily circulate our survey in the specialized Facebook groups where professional software developers (especially Java developers) discuss their programming problems and share software development resources. We also use LinkedIn\footnotemark[7]\footnotetext[7]{https://www.linkedin.com/} as a research tool to reach potential participants because it is one of the largest professional social networks in the world. We get 15 responses from this open circular, where six of them were incomplete. We thus consider nine valid responses to further analysis.
    
% \end{itemize}

\subsubsection{Survey Data Analysis}

In total, we received 53 valid survey responses. We then analyze the responses with appropriate tools and techniques based on the question types. For multiple-choice questions, we report the percentage of each option selected. To identify the agreement level of each statement, we analyze the Likert-scale ratings. We use Borda count \citep{Yamashita-DoDevelopersCareAboutCodeSmell-WCRE2013} to rank the reproducibility challenges. Furthermore, we extract comments that our survey participants give on the impacts of the reproducibility challenges and tool design requirements.

\begin{table*}[!htb]
\centering
% 	\captionsetup{justification=centering, labelsep=newline}
	\caption{Developers agreement to the reproducibility status and the challenges that prevent reproducibility}
% 	\vspace{-3mm}
	\label{table:results-agreement-analysis}
 	\resizebox{6.5in}{!}{%
    \begin{tabular}{l|c|c|c|c|c} \toprule
    \multirow{2}{*}{\textbf{Question No.}} & \multicolumn{2}{c|}{\textbf{Reproducibility Status}} & \multirow{2}{*}{\textbf{Reproducibility Challenges}} & \multicolumn{2}{c}{\textbf{Reproducibility Challenges}} \\ 
                       & \textbf{Irreproducible} & \textbf{Reproducible} &                 & \hspace{7mm}\textbf{Agree}\hspace{7mm} & \textbf{Disagree}    \\    \midrule
                     \multirow{2}{*} {01 (see Fig. \ref{fig:reproducibility-challenge-1})}  &  \multirow{2}{*} {94.3\%} &   \multirow{2}{*} {5.7\%}  &  \multicolumn{1}{l|}{$\bullet$ Class/Interface/Method not found}  &  92\%   &  8\%    \\
                                                                                            &                                 &                       & \multicolumn{1}{l|}{$\bullet$ Important part of code missing}     &  92\%   &  8\%  \\ \midrule
                     
                     \multirow{3}{*} {02 (see Fig. \ref{fig:reproducibility-challenge-2})}  &  \multirow{3}{*} {92.5\%}       &  \multirow{3}{*} {7.5\%}  &  \multicolumn{1}{l|}{$\bullet$ External library not found}  &  93.9\%   &  6.1\%   \\
                                                                                            &                                 &                       & \multicolumn{1}{l|}{$\bullet$ Identifier/Object type not found}     &   89.8\%   & 10.2\%    \\ 
                                                                                            &                                 &                       & \multicolumn{1}{l|}{$\bullet$ Too short code snippet}     &    93.9\%    & 6.1\%  \\ \midrule

                     \multirow{2}{*} {03 (see Fig. \ref{fig:reproducibility-challenge-3})} &  \multirow{2}{*} {92.5\%}        &   \multirow{2}{*} {7.5\%}  &  \multicolumn{1}{l|}{$\bullet$ Class/Interface/Method not found}  &  81.6\%   &  18.4\%  \\
                                                                                           &                                  &                            & \multicolumn{1}{l|}{$\bullet$ Database/File/UI dependency}      &  89.8\%   & 10.2\%    \\ \midrule

                     04 (see Fig. \ref{fig:reproducibility-challenge-4})                   &               73.6\%              &          26.4\%          &   \multicolumn{1}{l|}{$\bullet$ Outdated code} &    79.5\%     & 20.5\%    \\ \midrule
                    \textbf{ Overall}                                                      &               \textbf{ 88.2\%}    &    \textbf{11.8\%}        &                                --           &    \textbf{89.1\%}     &  \textbf{10.9\%}  \\ \bottomrule 
                     
    \end{tabular}
    }
    \vspace{-2mm}
\end{table*}

% \begin{itemize}
%     \item \textbf{RQ1 (part I), RQ2, and RQ3 (parts I, II).} We will present a set of closed-ended questions that limit participants to a list of predefined answer choices. Thus, the results will be described detailing the minimum, maximum, average, and median values. To test for significant differences among the responses, we perform statistical analysis using the Mann Whitney U test and Cliff’s delta. RQ3 (Part I) responses (i.e., sort the challenges from most harmful to least harmful), we will use the Borda count as proposed by 
% Yamashita and Moonen~\citep{Yamashita-DoDevelopersCareAboutCodeSmell-WCRE2013}.
% Borda count is a rank-order aggregation technique for $n$ candidates to rank. 
% The first ranked candidates received $n$ points, the second $n-1$ points, and so on. Similar approach will be applied to RQ3 (Part II), i.e., list top three challenges that need to be fixed first.
    
%     \item \textbf{RQ1 (parts II, III), and RQ4.} These research questions will be evaluated by some open-ended questions. Thus, we plan to use a card sorting approach \citep{zimmermann2016card} since we do not know the participants' responses in advance. Card sorting can infer themes from responses.
% \end{itemize}

\section{Survey Results}
\label{sec:survey-results}
In this section, we discuss the survey results and answer our research questions.

\subsection{Agreement Analysis to the Catalogue of Reproducibility Challenges (RQ1)} \label{agree-analysis-rq1}

We attempt to find developers' agreement on the challenges (see Table \ref{table:reproducibility-challenges}) that prevent reproducibility. Such agreement on empirical findings increases confidence in the findings \citep{rahman2020why}.

\noindent\textbf{Approach.} We present four SO questions (see Fig. \ref{fig:reproducibility-challenge-1}, \ref{fig:reproducibility-challenge-2}, \ref{fig:reproducibility-challenge-3} \& \ref{fig:reproducibility-challenge-4}) from the published dataset by Mondal et al. \citep{Mondal-SOIssueReproducability-MSR2019}. We first ask the participants to reproduce the question issues using the code segments included with the questions. We then ask either they can reproduce the issues using the code segments or not. Upon selection, ``I cannot reproduce'', we present a list of reproducibility challenges that could prevent reproducibility of the issues. We then ask their agreement/disagreement on those reproducibility challenges. Besides, we offer an option to mention if they find additional challenges. On the contrary, upon selection, ``I can reproduce'', participants are asked to mention their agreement/disagreement with a list of editing actions. However, we offer an option to mention if they perform additional actions.

Our primary focus is to validate the challenge catalog. Thus, we select four questions whose issues could not be reproduced by Mondal et al.~\citep{Mondal-SOIssueReproducability-MSR2019} and cover all the seven challenges (e.g., Table \ref{table:reproducibility-challenges}). Here, we ask the following three questions and analyze the participants' agreement in contrast to the reproducibility status (i.e., reproducible/irreproducible) and challenges reported by Mondal et al.~\citep{Mondal-SOIssueReproducability-MSR2019}. 
% We present four SO questions from the dataset of Mondal et al. \citep{Mondal-SOIssueReproducability-MSR2019}, whose issues could not be reproduced by them. 
% %We select these four questions because they cover all the reproducibility challenges shown in Table \ref{table:reproducibility-challenges}. 
% First, we attempt to see the developers' acceptance of the reproducibility status. We thus ask them whether they can reproduce the issues using the code segments included with the questions. We present the reproducibility challenges that prevent the reproducibility of the issues. Developers find those reproducibility challenges if they fail to reproduce the issues. We then ask the developers' agreement (yes/no) with the reproducibility challenges. Besides, they are requested to mention additional challenges if they find reproducing the issues. In particular, we ask three questions to the developers as follows. 

\begin{tcolorbox}[colframe=black!70,colback=white,left=0pt,right=1pt,top=1pt,bottom=1pt,boxrule=1pt,arc=1pt]
    \textbf{Q\textsubscript{1})} Do the participants reproduce the reported issues? (\emph{I can reproduce the issue/I cannot reproduce the issue})
    
    \textbf{Q\textsubscript{2})} Do the participants agree with the reproducibility challenges? (\emph{Yes/No})
    
    \textbf{Q\textsubscript{3})} Do the participants find additional challenges reproducing the issues? (\emph{Text})
\end{tcolorbox}

\noindent\textbf{Findings.} Table \ref{table:results-agreement-analysis} shows the summary of the findings. For question 1 (Fig. \ref{fig:reproducibility-challenge-1}), 94.3\% of the participants could not reproduce the issue. Only 5.7\% of them report that they could reproduce the issue. We found two reproducibility challenges for this code example that prevent reproducibility. They are - \emph{Class/Interface/Method not found} and \emph{Important part of code missing}. We see that 92\% of participants also encounter those two challenges. However, only 8\% of them disagree with those challenges. We see a similar agreement/disagreement to the reproducibility status and their associated challenges for the questions 2 (Fig. \ref{fig:reproducibility-challenge-2}) \& 3 (Fig. \ref{fig:reproducibility-challenge-3}). However, in the case of question 4 (Fig. \ref{fig:reproducibility-challenge-4}), about 26\% of developers can reproduce the reported issue, where the challenge was \emph{outdated code}. The code segment contains classes (e.g., \texttt{AudioStream}) that are not being used in JDK-1.8. However, some developers could use previous JDK versions to analyze the code or use alternative classes to reproduce the issue. We ask participants to submit the modified code segments and find that several participants compose new code to reproduce the issue.

In addition to the given challenges, participants report a few additional challenges they encountered as follows.

\begin{inparaenum}[(1)]
    \item \textbf{Error log/stack trace missing.} Error log or stack trace contains meaningful insights about program failures. Some issues could not be reproduced without such error reports.
    
    \item \textbf{System dependency or environment setup is missing.} Several programming issues are involved with particular Operating Systems (OS) (e.g., Windows), IDE (e.g., Eclipse), and software versions (e.g., Python 3). Some issues could not be reproduced because question submitters do not specify them.
    
    \item \textbf{Sample input-output missing.} Several issues demand sample input-output (i.e., test cases) while reproducing them.
\end{inparaenum}

\begin{table}[!htb]
	\centering
	\captionsetup{justification=centering, labelsep=newline}
	\caption{Agreement analysis by experience and profession \\  \footnotesize{(\textbf{SD:} Software Developer, \textbf{TL:} Technical Lead, \textbf{RE:} Research Engineer, \textbf{AP:} Academic Practitioner)}}
	\label{table:agreement-experience-profession}
	\resizebox{3.3in}{!}{%
    \begin{tabular}{l|c|c|c|c} \toprule
    \multicolumn{5}{c}{\textbf{Analysis by Java experience}} \\ \midrule
    \multirow{2}{*}{\textbf{Experience}} & \multicolumn{2}{c|}{\textbf{Reproducibility Status}} & \multicolumn{2}{c}{\textbf{Reproducibility Challenges}} \\ 
                                    & \textbf{Irreproducible}   & \textbf{Reproducible} & \textbf{Agree}   & \textbf{Disagree}\\ \midrule
    $\leq$ 5 years & 91.2\%  & 8.8\% & 85.1\%  & 14.9\%   \\ \midrule
    $>$ 5 years & 82.9\%  & 17.1\% & 92.2\%  & 7.8\%  \\ \midrule
    
   \multicolumn{5}{c}{\textbf{Analysis by profession}} \\ \midrule
   \multirow{2}{*}{\textbf{Profession}} & \multicolumn{2}{c|}{\textbf{Reproducibility Status}} & \multicolumn{2}{c}{\textbf{Reproducibility Challenges}} \\ 
                                    & \textbf{Irreproducible}   & \textbf{Reproducible}   & \textbf{Agree}   & \textbf{Disagree} \\ \midrule
    SD      & 86.5\%  & 13.5\%  & 89.8\%  & 10.2\% \\ \midrule
    AP  & 98.8\%  & 1.2\%  & 85.6\%  & 14.4\% \\ \midrule
    TL \& RE & 69.5\%  & 30.6\%  & 75.8\%  &  24.2\% \\ \bottomrule
    \end{tabular}
    }
    %\vspace{-.35cm}
\end{table}

We then analyze the agreement according to the participants' Java experience and profession. Table \ref{table:agreement-experience-profession} summarizes the agreement. We first measure the agreement with reproducibility status and challenges for each of the four example questions and then compute their average. We see that participants with lower experience (e.g., five years or less) agree 8\% more on average with the reproducibility status than those having higher experience (e.g., more than five years). Conversely, participants with higher experience agree 7\% more with the given reproducibility challenges. Participants with higher experience might have the skill to apply different potential approaches to reproduce the issues. However, low-experienced participants are more enthusiastic about resolving the challenges (e.g., finding external libraries, fixing database dependencies). Analysis by profession shows that technical leads and research engineers agree comparatively less with reproducibility status and challenges. They might not be actively involved in programming and have a hectic task schedule. Thus, they could guess the reproducibility status and the potential challenges. However, our investigation finds that they mainly disagree with the reproducibility status and challenge of question 4 (see Fig. \ref{fig:reproducibility-challenge-4}), where the challenge was ``outdated code''. 

\noindent\textbf{Summary.} About 88\% of participants (on average) fail to reproduce the issues using the code segments, and 89\% of them encounter the given challenges. Only about 11\% of participants (on average) claim that they could reproduce the issues, and thus they do not agree to the challenges. According to the agreement/disagreement analysis, at least 77\% more participants agree to the reproducibility status and challenges on average. Furthermore, our analysis by experience and profession also shows that the lowest agreement level with reproducibility status and challenges is 70\% (on average) which is acceptable. Such agreements support and validate the reproducibility status and challenge catalog (Table \ref{table:reproducibility-challenges}).

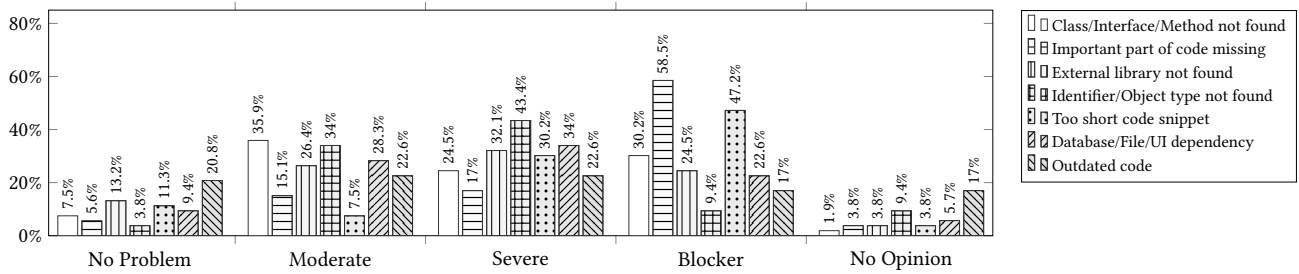
\begin{figure*}[!htb]
\centering
   	\pgfplotstableread{
    		1	7.5   5.6   13.2  3.8   11.3  9.4   20.8
    		2	35.9  15.1  26.4  34    7.5   28.3  22.6
    		3	24.5  17    32.1  43.4  30.2  34    22.6
    		4	30.2  58.5  24.5  9.4   47.2  22.6  17
    		5	1.9   3.8   3.8   9.4   3.8   5.7   17

    	}\datatable
    	
      \resizebox{6.8in}{!}{%
      \begin{tikzpicture}
        	\begin{axis}[
        	xtick=data,
        	xticklabels={No Problem, Moderate, Severe, Blocker, No Opinion
        	},
        	xticklabel style={font=\large},
        	enlarge y limits=false,
        	enlarge x limits=0.12,
        %	nodes near coords,
        	ymin=0,ymax=85,
        	ybar,
        	bar width=0.33cm,
        	width=6.8in,
        	height = 2.1in,
        	ytick={0,20,...,80},
            yticklabels={0\%,20\%,40\%,60\%,80\%},
        	ymajorgrids=false,
        %	xminorgrids=true,
        	yticklabel style={font=\large},
        	major x tick style = {opacity=0},
        	minor x tick num = 1,    
        	minor tick length=1ex,
        	legend style={
        	font=\small,
        	legend pos = outer north east,
        % 	legend pos=north west,
        %   anchor=west,
        % 	cells={align=left},
        	legend cell align=left
            },
            nodes near coords style={rotate=90,  anchor=west, font=\small},
        	nodes near coords =\pgfmathprintnumber{\pgfplotspointmeta}\%
        	%nodes near coords*={\pgfmathprintnumber[precision=2]\pgfplotspointmeta \%}
        	]
        	\addplot[draw=black!80, fill=black!0] table[x index=0,y index=1] \datatable;
        	\addplot[draw=black!80, fill=black!2, postaction={pattern=horizontal lines}] table[x index=0,y index=2] \datatable;
        	\addplot[draw=black!80, fill=black!4, postaction={pattern=vertical lines}] table[x index=0,y index=3] \datatable;
        	\addplot[draw=black!80, fill=black!6, postaction={pattern=grid}] table[x index=0,y index=4] \datatable;
        	\addplot[draw=black!80, fill=black!8, postaction={pattern=dots}] table[x index=0,y index=5] \datatable;
        	\addplot[draw=black!80, fill=black!10, postaction={pattern=north east lines}] table[x index=0,y index=6] \datatable;
        	\addplot[draw=black!80, fill=black!12, postaction={pattern=north west lines}] table[x index=0,y index=7] \datatable;
            \legend	{Class/Interface/Method not found,
            		 Important part of code missing,
            		 External library not found,
            		 Identifier/Object type not found,
            		 Too short code snippet,
            		 Database/File/UI dependency,
            		 Outdated code
            		 }
        	\end{axis}
    	\end{tikzpicture}
    	}
    %\vspace{-3mm}
\caption{Impacts on the reproducibility challenges answering questions.}
%\vspace{-1mm}
\label{fig:impacts-reproducibility-challenges}
\end{figure*}

\subsection{Impacts of Each of the Reproducibility Challenges to Answer Questions (RQ2)}
\label{impact-analysis}

We see the practitioners' agreement to the reproducibility challenge catalog in the previous section (Section \ref{agree-analysis-rq1}). However, it is important to understand the impact of each challenge to answer a question. Such understanding helps us to determine which challenges need to be addressed early.

\noindent\textbf{Approach.} To find the potential impact of the reproducibility challenges, we ask the participants to mark each of the challenges with one of the five categories: (1) not a problem, (2) moderate, (3) severe, (4) blocker, and (5) no opinion. Here, moderate means irritating, but able to answer appropriately. Severe means wasted much time to reproduce the issue, but able to answer, and blocker means could not reproduce the issue and thus could not answer the question. In particular, we ask a question to the participants as follows.

% \begin{tcolorbox}[flushleft upper,boxrule=1pt,arc=0pt,left=0pt,right=0pt,top=0pt,bottom=0pt,colback=white,after=\ignorespacesafterend\par\noindent]
\begin{tcolorbox}[colframe=black!70,colback=white,left=0pt,right=1pt,top=1pt,bottom=1pt, boxrule=1pt,arc=1pt]
    What are the impacts of the reproducibility challenges to answer a question? (\emph{not a problem/moderate/severe/blocker/no opinion})
\end{tcolorbox}

\noindent\textbf{Findings.} Fig. \ref{fig:impacts-reproducibility-challenges} shows the summary of how the participants assess the impact of each of the reproducibility challenges. We see that majority of the participants (about 36\%) perceive the challenge ``Class/Interface/Method not found'' as moderate. About 30\% of them consider it a blocker. However, only 7.5\% of participants consider this challenge not a problem, and the remaining 1.9\% of them did not give any opinion. ``An important part of code missing'' is mostly (about 59\% of participants) considered a blocker.  We find that ``Too short code snippet'' is also perceived as a blocker by the majority of the participants (about 47\%). About 43\% of participants perceive the challenge ``Identifier/Object type not found'' as severe, whereas 34\% of them perceive it as moderate. 

We then attempt to find the top challenges in each of the five categories. As Fig. \ref{fig:impacts-reproducibility-challenges} shows, ``Class/Interface/Method not found'' is the top in the moderate challenge category. The challenge ``Identifier/Object type not found'' is the top in the severe category, whereas ``Important part of code missing'' is assessed as the top in blocker category. However, among the reproducibility challenges, ``Outdated code'' is considered ``not a problem'' by most participants, or many of them did not give any opinion. To reproduce the issue, developers could migrate the outdated code or use alternative classes/APIs in place of obsolete classes/APIs. Thus, this challenge could be estimated mostly as not a problem among the seven challenges.

In addition to that, we attempt to get more insights into participants' choices of impact. We thus ask their justifications behind the choices. Although it was optional, we receive a few of their comments (see Table \ref{table:developers-comments-severity}). For example, one developer states that without the Class/Method definition, it is nearly impossible to reproduce the exact issues.

\noindent\textbf{Summary.} Most of the developers could not reproduce the question issues and thus could not answer if code segments miss important statements. Other challenges (e.g., Class/Interface/Method and Identifier/Object type not found) could irritate developers and kill their valuable time. Such factors could also prevent or delay appropriate solutions to the questions. However, developers do not encounter more difficulties in dealing with outdated code.

\begin{table*}[!htb]
\centering
% 	\captionsetup{justification=centering, labelsep=newline}
	\caption{Practitioners' comments behind the choices of impacts to answer questions}
% 	\vspace{-3mm}
	\label{table:developers-comments-severity}
 	\resizebox{6.8in}{!}{%
    %\rowcolors{1}{}{lightgray}
    \begin{tabular}{p{16cm}} \toprule
    % \multicolumn{1}{c}{\textbf{Comments from Developers}} \\ \midrule
    \emph{``I really think not finding `Class/Interface/Method' is a blocker problem, because how can I understand a code without their method explained and reproduce. Same case happens in the case of  `Too short code snippet' and 'Important part of code missing'.''} \\ %\arrayrulecolor{black!0} \midrule
    
    \emph{``Without Class/Method definition, it's nearly impossible to identify the exact error. Besides, it is helpful, if individuals give some sample input and output examples. Dependency on Database: this is kind of blocker to run the sample code. ''} \\ %\arrayrulecolor{black!0} \midrule

    \emph{``If the given code part is too small, solution can not be made by depending only few line of code.''} \\ %\arrayrulecolor{black!0} \midrule
    
    \emph{``As I already answered most of the time we are irritated to help with question without proper playground. ( class not found, Important code missing... etc.).''} \\ \bottomrule %\arrayrulecolor{black}
\end{tabular}
}
\end{table*}

\subsection{Severity Analysis of the Reproducibility Challenges (RQ3)}

It is not always possible to fix all the reproducibility challenges due to a strict budget (e.g., limited time). Thus, it is important to prioritize more severe challenges than the others to fix them within a strict budget.

\noindent\textbf{Approach.} We ask the following question to participants to find the rank of the reproducibility challenges.

\begin{tcolorbox}[colframe=black!70,colback=white,left=0pt,right=1pt,top=1pt,bottom=1pt,boxrule=1pt,arc=1pt]
    Which three challenges the participants would like to prioritize above others? (\emph{first choice/second choice/third choice})
\end{tcolorbox}

\noindent We then use the Borda count as proposed by Yamashita and Moonen~\citep{Yamashita-DoDevelopersCareAboutCodeSmell-WCRE2013} to analyze and rank the challenges according to their severity. Borda count is a rank-order aggregation technique for $n$ candidates. The first ranked candidates received $n$ points, the second $n-1$ points, and so on. Since we ask three options to choose,  we assign the first option to $3$ points, the second to $2$ points, and the third to $1$ point. 

\begin{figure}[!htb]
\centering
        \resizebox{3in}{!}{%
    	\begin{tikzpicture}
        	\begin{axis}
            [
            height = 4.8cm,
            width = 2in,
            xmin=0, xmax=130,
            xbar,
            ytick = data,
            enlarge y limits=0.15,
        	enlarge x limits=false,
	        nodes near coords,
            % xlabel = Score,
            % y axis line style = { opacity = 0 },
            % axis x line       = none,
            % tickwidth         = 0pt,
            % enlarge y limits  = 0.01,
            % enlarge x limits  = 0.01,
            bar width=0.28cm,
            symbolic y coords = {
            Important part of code missing,
            Class/Interface/Method not found,
            Too short code snippet,
            External library not found,
            Database/File/UI dependency,
            Identifier/Object type not found,
            Outdated code
            },
            % nodes near coords,
            legend style={at={(0.5,-0.20)},
        	font=\footnotesize,
            anchor=north,legend columns=-1}
            ]
            \addplot[xbar,fill=black!50] coordinates {
            (101,Important part of code missing)
            (87,Class/Interface/Method not found)
            (57,Too short code snippet)
            (39,External library not found) 
            (16,Database/File/UI dependency) 
            (13,Identifier/Object type not found) 
            (5,Outdated code) 
            };
           \end{axis}
        \end{tikzpicture}
        }
% \vspace{-2mm}
\caption{The rank of reproducibility challenges according to their score.}
% \vspace{-3mm}
\label{fig:severity-reproducibility-challenges}
\end{figure}
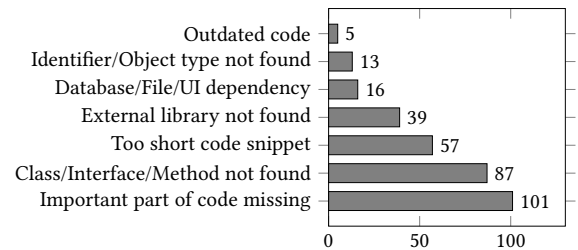

\begin{table*}[!htb]
\centering
% 	\captionsetup{justification=centering, labelsep=newline}
	\caption{Editing actions to make the code segments capable of reproducing the question issues.}
% 	\vspace{-3mm}
	\label{table:editing-actions}
 	\resizebox{6.5in}{!}{%
    \begin{tabular}{p{6em}|p{16em}|p{20em}|c} \toprule
    
    \textbf{Question No.}        &  \multicolumn{1}{c|}{\textbf{Reproducibility Challenges}} & \multicolumn{1}{c|}{\textbf{Action Detail}} & \textbf{No Action}  \\ \midrule
    
    01 (see Fig. \ref{fig:reproducibility-challenge-1})& (1) Class/Interface/Method not found, and (2) Important part of code missing & Addition of demo classes and methods (62\%), object creation, identifier declaration and initialization (20\%), invocation of methods (4\%) & 36\% \\ \midrule
    
    02 (see Fig. \ref{fig:reproducibility-challenge-2})& (1) External library not found, (2) Identifier/Object type not found, and (3) Too short code snippet & Addition of demo classes and methods (26.5\%), inclusion of native and external libraries (14.3\%), object creation, identifier declaration and initialization (12.2\%), other (8.2\%)   & 61.2\% \\ \midrule
    
    03 (see Fig. \ref{fig:reproducibility-challenge-3})& (1) Class/Interface/Method not found, and (2) Database/File/UI dependency & Addition of demo classes and methods (28\%), inclusion of native libraries (12\%), other (4\%)  & 68\%  \\ \midrule
    
    04 (see Fig. \ref{fig:reproducibility-challenge-4})& Outdated code & Addition of demo classes and methods (20.5\%), inclusion of native and external libraries (20.5\%), suggest new code (10.3\%), other (5.1\%)  & 53.8\% \\ \bottomrule 
                     
    \end{tabular}
    }
\end{table*}

\noindent\textbf{Findings.} Fig. \ref{fig:severity-reproducibility-challenges}  shows the challenges with their points. We see that that ``important part of code missing'' gets the highest (i.e., 101) and ``Outdated code'' get the lowest (i.e., 5) point. However, we can now prioritize (e.g., rank) the top three challenges according to their severity --  (1) important part of code missing, (2) class/interface/method not found, and (3) too short code snippet. This finding of severity analysis is consistent with impact analysis (section \ref{impact-analysis}).

\noindent\textbf{Summary.} The findings from severity analysis suggest that question submitters should include the part of code (i.e., statements) that could never be guessed, define necessary classes/methods. Moreover, they should not submit too short code segments (e.g., 1/2 lines of code) to receive fast and appropriate solutions.

\subsection{Editing Actions to Make the Code Segments Capable to Reproduce the Issues (RQ4)}

We could get more insights on how to address the reproducibility challenges by observing the developers' code modification plan.

\noindent\textbf{Approach.} We attempt to see the participants' code editing actions to reproduce the question issues. In particular, we want to see how participants mitigate the reproducibility challenges of the code segments. We thus ask participants to submit their modified code segments and manually level the editing actions. We shared the modified code with the level of actions in our online appendix \citep{ourdataset}.

\noindent\textbf{Findings.} Table \ref{table:editing-actions} shows the summary of the code editing actions to make the code segments capable of reproducing the issues. The participants add demo classes and methods, include import statements for the native and external libraries, create objects, declare identifiers and initialize them, and migrate the outdated code to mitigate the reproducibility challenges. In particular, participants first attempt to make the code segments compilable/executable by resolving the external dependencies. For example, when methods are invoked without defining them, participants attempt to add their definitions. Participants modify the codes based on the given code segments, hints from question descriptions or guessing. However, we see that 36\% -- 68\% of participants did not modify the code segments. Intuitively, they assume that the code segment is insufficient to reproduce the issues, and thus they did not attempt to modify the code segments. For example, one developer commented, \emph{``I did not try to reproduce it, because it needs much guessing and it may differ much from the original code''}.	

\begin{table*}[htb]
\centering
% 	\captionsetup{justification=centering, labelsep=newline}
	\caption{Practitioners' tool support recommendations to promote reproducibility}
% 	\vspace{-3mm}
	\label{table:tool-support-recommendations}
 	\resizebox{6.5in}{!}{%
    %\rowcolors{1}{}{lightgray}
    \begin{tabular}{p{18cm}} \toprule
    % \multicolumn{1}{c}{\textbf{Tool Support Recommendations}} \\ 
    \emph{``The tool that help me to guess the missing code, library version, missing input data depending on the issue context. It will be great if the tool generates missing information for me depending on the question/issue.''} \\  %\arrayrulecolor{black!0} 
    
    \emph{``A software that can ensure the missing cases and force the user to complete them, Auto code generator.''} \\  %\arrayrulecolor{black!0} 

    \emph{``A tool that helps to guess the missing important code depending on the question context.''} \\ %\arrayrulecolor{black!0} 
    
    \emph{``A tool that can sense my written code snippet and see what important part  might be missing there.''} \\ %\arrayrulecolor{black!0} 
    
    \emph{``It will provide suggestions to reproduce after pasting the code in IDE. ''} \\ \bottomrule %\\ \arrayrulecolor{black} 

\end{tabular}
}
% \vspace{-.2in}
\end{table*}

\noindent\textbf{Summary.} Developers' first attempt to make the code segments compilable/executable by resolving all possible dependencies. However, developers do not modify the code segments based on guessing when the code segments are insufficient.

\begin{table*}[htb]
\centering
% 	\captionsetup{justification=centering, labelsep=newline}
	\caption{Assessment of the tool support options}
% 	\vspace{-3mm}
	\label{table:tool-support-options}
 	\resizebox{6.5in}{!}{%
    %\rowcolors{1}{}{lightgray}
    \begin{tabular}{p{12cm}|c|l} \toprule
    
    \multicolumn{1}{c|}{\textbf{Options}} & \multicolumn{1}{c|}{\textbf{Mean Value}} & \multicolumn{1}{c}{\textbf{Result Interpretation}} \\ \midrule
    A tool that suggests users improve the code examples to support reproducibility                             &   4.43 &  Very Influential    \\ \midrule
    A tool that warns users about reproducibility challenges that are severe or may block the reproducibility   &   4.25 &  Very Influential    \\ \midrule
    An IDE (e.g., Eclipse) plugin to support reproducibility                                                    &   4.25 &  Very Influential    \\ \midrule
    A browser plugin to support reproducibility                                                                 &   4.02 &  Influential    \\ \midrule
    A website that guides users to improve reproducibility                                                      &   3.85 &  Influential    \\ \midrule
    A static analyzer that examines code segments to find reproducibility challenges                            &   4.15 &  Influential    \\ \bottomrule 
\end{tabular}
}
\vspace{-2mm}
\end{table*}

\subsection{Tool Support Needs to Promote Reproducibility (RQ5)}

Questions whose issues could be reproduced have at least three times higher chance to receive acceptable answers than the questions with irreproducible issues. However, the existing question submission system of SO is not intelligent enough to estimate reproducibility. It does not interact with the question submitters even if the code segments fail to reproduce the question issues. To enhance the possibility of receiving an appropriate answer and save developers' valuable time, we plan to introduce interactive tool support (e.g., browser/IDE plugin) to promote reproducibility.

\noindent\textbf{Approach.} We seek participants' recommendations on tool design requirements. Besides, we offer a few tool support options (Table \ref{table:tool-support-options}) and employ a 5-point Likert scale (i.e., 1--5) to estimate the participants' consent with the tool options. In particular, we ask two questions as follows.

\begin{tcolorbox}[colframe=black!70, colback=white,left=0pt,right=1pt,top=1pt,bottom=1pt,boxrule=1pt,arc=1pt]
    \textbf{Q\textsubscript{1})} What kind of tool support do the participants need to assist with the reproducibility challenges? (\emph{Text})
    
   \textbf{Q\textsubscript{2})} How do the participants agree with our tool support options? (\emph{see options from Table \ref{table:tool-support-options}})
\end{tcolorbox}

\noindent\textbf{Findings.}
Table \ref{table:tool-support-recommendations} shows a few valuable recommendations from participants (complete recommendation list can be found in our online appendix \citep{ourdataset}). 
% Most of them recommend introducing a tool that can analyze the code segments and suggest question submitters improve the code segments with the missing part of code, library inclusion, classes/methods definitions, and sample input-output. 
For example, one participant recommended, \emph{``A tool that can sense my written code snippet and see what important part might be missing there''}.
Besides, Table \ref{table:tool-support-options} shows our tool options. We offer six options, such as a tool that warns users about the severe challenges or challenges that could block reproducibility. According to the participants' assessment, three of them are considered very influential (score $\geqslant$ 4.21), and the remaining three are considered influential (3.41 $\leqslant$ score $\leqslant$ 4.20). 

% \begin{figure}[htb]
%   \centering
%   \includegraphics[width=2.7in]{tlx-plot.png}
%   \caption{\small{NASA TLX scores to evaluate the cognitive workload to reproduce the question issues (\textbf{MT}: Mental, \textbf{PC}: Physical,  \textbf{TR}: Temporal,  \textbf{PM}: Performance,  \textbf{EF}: Effort,  \textbf{FT}: Frustration).}}
%   \label{fig:nasa-tlx-score}
% \end{figure}

\noindent\textbf{Summary.} By analyzing the developers' recommendations and assessing the given tool options, we could summarize the primary tool design requirements as follows. 

\begin{inparaenum}

    \item[$\bullet$] Interact with question submitters and suggest including the part of code that could never be guessed and mandatory to reproduce the question issues.
    
    \item[$\bullet$] Suggest adding the definition of necessary classes/methods and declaration of identifiers/objects. The absence of such definitions/declarations could prevent reproducibility.
    
    \item[$\bullet$] Warn question submitters not to include a too short code segment. Developers could not guess the actual problem from such a code segment. Sometimes developers guess the problem. However, such guessing often goes wrong, and thus the submitted answer may not solve the actual problem.
    
    \item[$\bullet$] Recommend to include the external libraries when required. It is a time-consuming task to find the external libraries without import statements or any hints about the libraries in the question description.
    
\end{inparaenum}

\section{Threats to Validity}
\label{threattovalidity}

Threats to internal validity relate to experimental errors and biases \citep{tian2014automated}. Our key reproducibility challenges were derived from a qualitative study by Mondal et al. \citep{Mondal-SOIssueReproducability-MSR2019}, which could be a source of subjective bias. However, the challenges were validated by 53 developers with an agreement level of about 90\% on average. Moreover, the difference between agreement and disagreement on reproducibility challenges is quite large (about 77\%) and statistically significant.

Threats to external validity relate to the generalizability of our findings \citep{tian2014automated}. We investigate the practitioners' perspective on reproducibility challenges, their impacts and tool design requirements. However, the challenges were derived by analyzing questions related to Java programming problems. Thus, the impacts and tool design requirements can apply to the statically-typed, compiled programming languages such as C++ and C\#. Nonetheless, replication of our study using different languages (e.g., dynamically-typed languages like Python) may prove fruitful.

Our survey participants range from novice to experienced (see Fig. \ref{fig:java-experience}) and constitute mainly software developers but also other related professions (see Fig. \ref{fig:profession}). Such diversity in the survey participants offers validity and applicability to the survey findings. However, any individual bias in the survey responses should be mitigated via a large sample of 53 users.

% Threats to \emph{internal validity} relate to experimental errors and
% biases \citep{tian2014automated, rahman2020why}. Our key challenges behind issues irreproducibility and issue reproducibility status were derived from a qualitative study by Mondal et al. \citep{Mondal-SOIssueReproducability-MSR2019}, which could be a source of subjective bias. However, a group of 53 developers validated the reproducibility challenges and status with an agreement level of about 90\% (RQ1). Hence, such a threat might be mitigated. 
% \section{Limitation}
% \label{limitation}
% In this study, we investigate the challenges of reproducibility that are related to statically typed programming languages (especially Java). Thus, the impacts and tool design requirements may differ for dynamically typed programming languages (e.g., Python). We believe some challenges are common (e.g., missing library information). However, there might be additional challenges that demand further investigation.

\section{Conclusions}
\label{conclusion}

Developers submit thousands of questions to SO daily to resolve their code-level problems. They include an example code segment to support the problem descriptions. Users of SO prefer to reproduce the issues using the code segments and then submit their answers. However, such code segments could not always reproduce the issues due to several unmet challenges that prevent the questions from getting prompt and acceptable solutions. A previous study produces a catalog of challenges (Table \ref{table:reproducibility-challenges}) that might prevent reproducibility of question issues. However, they are not validated by developers. We survey 53 developers (users of SO) to understand their perspectives on those challenges, impacts of those challenges, how developers address those challenges, and tool design requirements that could mitigate those challenges. Our findings are fourfold. 
First, about 90\% of developers agree to the reproducibility challenge catalog. 
Second, missing part of code that could never be guessed but required to reproduce question issue mostly prevent questions from receiving answers. 
Third, developers perform several editing actions (e.g., addition of demo classes/methods) to make the code segments reproduce the issues. However, they do not try to reproduce the issues and submit answers if the code segments are entirely insufficient to guess any actions.
Finally, an intelligent tool that identifies the reproducibility challenges according to their severity could help question submitters improve their code segments and reduce the developers' workload who attempt to answer questions.

\vspace{1mm}
\textbf{Acknowledgement:} This research is supported by the Natural Sciences and Engineering Research Council of Canada (NSERC) and by two Canada First Research Excellence Fund (CFREF) grants coordinated by the Global Institute for Food Security (GIFS) and the Global Institute for Water Security (GIWS).

% \begin{acks}
% To do
% \end{acks}

\balance

% \begin{acks}
% This research is supported by the Natural Sciences and Engineering Research Council of Canada (NSERC), and by a Canada First Research Excellence Fund (CFREF) grant coordinated by the Global Institute for Food Security (GIFS).
% \end{acks}

% \balance
% \bibliographystyle{plainnat}
\bibliographystyle{ACM-Reference-Format}
\bibliography{bibliography}

\end{document}